\documentclass[twocolumn,showpacs,amsmath,amssymb]{revtex4}
\usepackage{graphicx}% Include figure files
\usepackage{dcolumn}% Align table columns on decimal point
\usepackage{bm}% bold math
\newcommand{\beq}{\begin{equation}}
\newcommand{\eeq}{\end{equation}}
\newcommand{\bea}{\begin{eqnarray}}
\newcommand{\eea}{\end{eqnarray}}
\newcommand{\eps}{\varepsilon}
\newcommand{\bfg}{\boldsymbol}

\begin{document}

\author{S. V. Tolokonnikov}
\affiliation{Kurchatov Institute, 123182 Moscow} \affiliation{Moscow
Institute of Physics and Technology, 123098 Moscow, Russia}
\author{S. Kamerdzhiev}
\affiliation{Institute for Physics and Power Engineering, 249033
Obninsk, Russia}
\author{S. Krewald}\affiliation{Institut f\"ur Kernphysik, Forschungszentrum J\"ulich,
D-52425 J\"ulich, Germany}
\author{E. E. Saperstein}
\affiliation{Kurchatov Institute, 123182 Moscow}
\author{D. Voytenkov}
\affiliation{Institute for Physics and Power Engineering, 249033
Obninsk, Russia}

\title{Effects of density dependence of the effective pairing interaction
on the first $2^+$ excitations and quadrupole moments of odd nuclei}

\pacs{21.10.-k, 21.10.Jx, 21.10.Re, 21.60-n}

\begin{abstract}
Excitation energies and transition probabilities of the first $2^+$
excitations in even tin and lead isotopes as well as  the quadrupole
moments  of odd neighbors of these isotopes are calculated within
the self-consistent Theory of Finite Fermi Systems based on the
Energy Density Functional by Fayans et al. The effect of the density
dependence of the effective pairing interaction is analyzed in
detail by comparing results obtained with  volume and surface
pairing. The effect is found to be noticeable. For example, the
$2^+$-energies are systematically higher at 200-300 keV for the
volume paring as compared with the surface  pairing case. But on the
average both models reasonably agree with the data. Quadrupole
moments of odd-neutron nuclei are very sensitive to the
single-particle energy of the state $\lambda$ under consideration
due to the Bogolyubov factor ($u^2_{\lambda}-v^2_{\lambda}$). A
reasonable agreement with experiment for the  quadrupole moments has
been obtained for the most part of odd nuclei considered. The method
 used gives a reliable  possibility to predict quadrupole
 moments of unstable odd nuclei including very neutron rich ones.

\end{abstract}

\maketitle

\section{Introduction}
Presently there are two theoretical approaches which can
quantitatively describe the bulk properties of nuclear isotope
chains with a small number of effective coupling constants:
selfconsistent mean field theories and density-functional theory.
The successes and open problems of the mean field approaches are
reviewed in Refs. \cite{REL-REV,SHF-REV1,SHF-REV2}. The Kohn-Sham
density functional theory was originally proposed for  chemistry and
solids \cite{KS,Jones}. Important theoretical developments have been
made: an extension of the Hohenberg-Kohn theorem to pairing degrees
 of freedom by Oliveira, Gross, and Kohn
 allowed studies of superfluids \cite{oliveira} and the generalization of functional theory
to study excited states made it possible to investigate the electromagnetic response
 of correlated electron materials \cite{basov}.
In nuclear physics, a self-consistent Theory of Finite Fermi Systems
(TFFS)
 was derived by Khodel and Saperstein \cite{KhS} on the basis of
the TFFS by  Migdal  \cite{AB} supplemented with the many-body
theory self-consistency relation \cite{Fay-Khod} for the nucleon
mass operator. As it was shown in \cite{Kh-Sap-Zv}, the
self-consistent TFFS for nuclei without pairing can be reformulated
as a particular version of the density functional method with a
rather complicated $\rho$-dependence of the energy functional. It
contains also $\tau$-dependent terms but with rather small strength
resulting for the effective mass in a small difference of
$|m^*_{n,p}(r)-m|\simeq 0.05m$. In a series of articles by Fayans et
al. \cite{STF,Fay5}, (see also \cite{Fay} and Refs. therein) the
  energy density functional (EDF) method
 was generalized for superfluid nuclei. Just as in the original Kohn--Sham approach,
the identity $m^*=m$ was imposed. A fractional form of the density
dependence for the central part of the normal component of the EDF
was introduced. The coordinate dependence of it resembled that of
\cite{Kh-Sap-Zv} but the functional form was much simpler making the
self-consistent QRPA calculations easier.
Note that a recent generalization of the Skyrme force
in \cite{EKR} contains a new term with a density dependence resembling
that in \cite{Fay}. In addition, the velocity dependent force in \cite{EKR}
is rather weak leading to the effective mass $m^*\simeq 0.9 m$. Thus, the
selfconsistent mean field approaches may eventually converge with the
density functional methods.

The non-relativistic versions of the self-consistent mean field
theories introduce three-nucleon forces which are often expressed as
a density dependent two-body interaction. In general, one assumes a
fractional power of the density dependence. Recent advances in
Effective Field Theory open the possibility to connect the density
functional with the effective two- and three- nucleon systems which
are determined from two-nucleon scattering and few-nucleon
reactions. Reviews about the current status of such attempts are
given in Refs.\cite{Epelbaum:2008ga,Drut:2009ce}.

The question arises whether the pairing interaction should have an
analogous dependence on the normal nuclear density. Several studies
derived pairing interactions from free two-nucleon interactions.
 Baldo et al. solved the gap equation in semi-infinite
nuclear matter \cite{Bald1}, nuclear slab \cite{Bald2}, and finite
nuclei \cite{Bald3,Bald4}.  The Paris and Argonne v$_{18}$ $NN$
potentials were used, results being almost identical. To make
results more appropriate for practical nuclear self-consistent
calculations dealing with pairing in a model space, the pairing
problem was treated in a two-step way. The gap equation was solved
in a model space with limiting energy $E_0= 30\div 40\;$MeV with the
use of the effective pairing interaction. The latter is found in the
subsidiary sub-space in terms of a free $NN$ potential. For all
systems under consideration and the two $NN$ potentials the
effective pairing interaction found is much stronger, up to ten
times, at the surface than inside. The Milan group concentrated on
the $^{124}$Sn nucleus, a traditional benchmark for the nuclear
pairing problem, and solved the gap equation starting from the
Argonne v$_{14}$ potential \cite{milan1}. In addition to the free
$NN$ interaction, they included corrections due to exchange with
low-lying surface vibrations (``phonons'') \cite{milan2} and
high-lying excitations, mainly spin-dependent ones, \cite{milan3}.
In the last article, a local 3-parameter density-dependent effective
pairing interaction is constructed for the model space with
$E_0=60\;$MeV which reproduce approximately  exact gap values.
Qualitatively, it is similar to that described above. Without all
corrections, it consists of a strong surface attraction and very
weak attraction inside. Taking into account of the phonon exchange
makes the inner interaction repulsive. At last, inclusion of the
spin-dependent excitations makes the inner repulsion rather strong.
Thus, the {\it ab initio} calculations of the effective pairing
interaction predict essential density dependence with strong surface
attraction.

As an alternative to consideration of the gap equation with complete
realistic $NN$ interactions, Bulgac and Yu used the fact that this
equation depends mainly on the low-$k$ behavior of $NN$ force which
can be approximated with a rather simple analytical function. It
helped to  develop  a renormalization scheme for the gap equation
without any cutoff in terms of zero-range interactions with explicit
coordinate dependence of the effective pairing interaction and to
suggest an EDF for superfluid nuclei \cite{Bulgac1,Bulgac2}.

The calculations by Fayans et al. employed both volume pairing and
surface pairing interactions. The binding energies and the proton
and neutron separation energies were found to be insensitive to the
type of pairing force used. But the odd-even staggering of charge
radii can be quantitatively reproduced only if the strong density
dependence of the pairing force is introduced \cite{Fay}.

In this work, we investigate the excitation energies and transition
rates of the  low-lying $2^+$-states in spherical nuclei with the
aim to analyze the sensitivity of those  observables  to the details
of the pairing interaction. We will compare two opposite limits, the
``volume pairing'' with density independent effective pairing
interaction ${\cal F}^{\xi}$ and the case of the function ${\cal
F}^{\xi}$ with the surface dominance. The latter will be named for
brevity the ``surface pairing''. Several sets of calculations of
these characteristics of the first $2^+$-excitations were carried
out recently within the QRPA method with Skyrme force
\cite{Bertsch1,Vor} and within the Generator Coordinate Method with
the Gogny force \cite{Bertsch2}. No systematic analysis of the
density dependence of the pairing force was performed in these
studies. Dealing with low-laying quadrupole excitations, it is
natural to include into analysis also quadrupole moments of odd
nuclei which give test of static quadrupole polarization.

In this paper, we use the EDF method \cite{Fay} with the functional
DF3-a \cite{Tol-Sap}. In the latter the spin-orbit and effective
tensor terms of the original functional DF3 \cite{Fay5,Fay} were
modified.  All the QRPA-like TFFS equations are solved in the
self-consistent basis $(\eps_{\lambda}, \varphi_{\lambda})$ obtained
within the EDF method with the functional DF3-a.

\section{Brief outline of the formalism}
For completeness, we describe shortly the EDF method of \cite{Fay}
using mainly the notation of \cite{Fay3}. In this method, the ground
state energy of a nucleus is considered as a functional of normal
and anomalous densities, \beq E_0=\int {\cal E}[\rho_n({\bf
r}),\rho_p({\bf r}),\nu_n({\bf r}),\nu_p({\bf r})] d^3r.\label{E0}
\eeq  The normal part of the EDF ${\cal E}_{\rm norm}$ contains the
central, spin-orbit and effective tensor nuclear terms and Coulomb
interaction term for protons. The main, central force, term of
${\cal E}_{\rm norm}$ is finite range with Yukawa-type coordinate
dependence. It is convenient to extract the $\delta({\bf r}-{\bf
r}')$-term from the Yukawa function separating the rest of \beq
D({\bf r}-{\bf r}')=\frac 1 {4\pi r_{\rm c}^2 |{\bf r}-{\bf r}'|}
\exp{\left( - \frac {|{\bf r}-{\bf r}'|} {r_{\rm c}} \right)}-
\delta({\bf r}-{\bf r}') \label{D_r}\eeq to generate the ``surface''
part ${\cal E}^{\rm s}$ which vanishes in infinite matter with
$\rho({\bf r})=const$. The Yukawa radius $r_{\rm c}$ is taken the
same for the isoscalar and isovector channels. The ``volume'' part
of the EDF, ${\cal E}^{\rm v}(\rho)$, is taken in
\cite{Fay5,Fay,Fay3} as a fractional function of densities
$\rho_+=\rho_n+\rho_p$ and $\rho_-=\rho_n-\rho_p$: \beq {\cal
E}^{\rm v}(\rho)=C_0 \left[ a^{\rm v}_+\frac{\rho_+^2}2 f^{\rm
v}_+(x) + a^{\rm v}_-\frac{\rho_-^2}2 f^{\rm v}_-(x)\right],
\label{EDF_v}\eeq where \beq f^{\rm v}_{\pm}(x)=\frac{1-h^{\rm
v}_{1\pm}x}{1+h^{\rm v}_{2\pm}x}. \label{fx_v}\eeq Here
$x=\rho_+/(2\rho_0)$ is the dimensionless nuclear density where
$\rho_0$ is the density of nucleons of one kind in equilibrium
symmetric nuclear matter. The factor $C_0=(dn/d\eps_{\rm F})^{-1}$
in Eq. (\ref{EDF_v}) is the usual TFFS normalization factor, inverse
density of states at the Fermi surface.

To write down the surface term in a compact form similar to (\ref{EDF_v}), the ``tilde'' operator
was introduced in \cite{Fay3} denoting the following folding procedure:
\beq
\widetilde{\phi({\bf r})}=\int D({\bf r}-{\bf r}')\phi({\bf r'}) d {\bf r'}.
\eeq

Then we obtain
\beq {\cal E}^{\rm s}(\rho)=C_0 \frac 1 2 \left[ a^{\rm s}_+ (\rho_+ f^{\rm s}_+)
\widetilde{(f^{\rm s}_+\rho_+)}  +
a^{\rm s}_- (\rho_- f^{\rm s}_-)\widetilde{(f^{\rm s}_-\rho_-)}\right],
\label{EDF_s}\eeq
where
\beq f^{\rm s}_{\pm}(x)=\frac{1}{1+h^{\rm s}_{\pm}x}.
\label{fx_s}\eeq
All the above parameters, $a^{\rm v}_{\pm}, a^{\rm s}_{\pm},
h^{\rm v}_{1\pm}h^{\rm v}_{2\pm},h^{\rm s}_{\pm}$, are dimensionless.

In the momentum space, the operator (\ref{D_r}) reads \beq
D(q)=-\frac {(q r_{\rm c})^2}{1+(q r_{\rm c})^2}. \label{D_q}\eeq In
the small $r_{\rm c}$ limit it reduces to $D(q)= -(q r_{\rm c})^2$,
and Eq. (\ref{EDF_s}) could be simplified to a Skyrme-like form
proportional to $(\nabla \rho)^2$.

The spin-orbit interaction reads
\beq
 {\cal F}_{sl}=C_0 r_0^2 (\kappa + \kappa' {\bfg\tau}_1{\bfg\tau}_2)
 \left[ \nabla_1 \delta ({\bf r}_1-{\bf r}_2) \times
 ({\bf p}_1-{\bf p}_2)\right]\cdot
({\bfg\sigma}_1 + {\bfg\sigma}_2), \label{Fsl}
\eeq
where the factor $r_0^2$ is introduced to make the spin-orbit
parameters  $\kappa, \kappa'$ dimensionless. It can be expressed in
terms of the above equilibrium density,
 $r_0^2=(3/(8\pi \rho_0))^{2/3}$.

In nuclei with partially occupied spin-orbit doublets, the so-called
 spin-orbit density exists, \beq
\rho_{sl}^{\tau}(\bf r)=\sum_{\lambda} n_{\lambda}^{\tau} \langle
\phi_{\lambda}^{\tau *}(\bf r) ({\bfg \sigma}{\bf l})
\phi_{\lambda}^{\tau}(\bf r)\rangle, \label{rhosl} \eeq where $\tau
=n,p$
--- is the isotopic index and averaging over spin variables is carried out.
As it is well known, see e.g. \cite{KhS}, a new term appears in the
spin-orbit mean field induced by  the tensor forces and the first
harmonic $\hat{g_1}$ of the spin Landau--Migdal (LM) amplitude. We
combine those contributions into an effective tensor force or first
spin harmonic, \beq {\cal F}_1^s=C_0r_0^2 (g_1 + g'_1
{\bfg\tau}_1{\bfg\tau}_2)
 \delta ({\bf r}_1-{\bf r}_2) ({\bfg\sigma}_1 {\bfg\sigma}_2)
 ({\bf p}_1{\bf p}_2). \label{g1}
\eeq

In Table 1, we present all parameters of the normal part of the EDF DF3-a
we use. Note that the major part of these parameters is identical to the
ones used in the DF3 functional \cite{Fay}.
With one exception, all
parameters for the central force part  remained the same
 and only the spin-orbit
and the first spin harmonic are changed according \cite{Tol-Sap}.
Application of the volume part (\ref{EDF_v}) to equilibrium nuclear
matter, with the equilibrium relation, i.e. vanishing pressure
$p(\rho)=\rho^2 \partial ({\cal E}/\rho)/\partial \rho$, permits to
find the parameters $a^{\rm v}_{+}, h^{\rm v}_{1+}$ and $ h^{\rm
v}_{2+}$ in terms of the nuclear matter density $\rho_0$, the
chemical potential $\mu_0$, and the compression modulus
$K_0=9dp/d\rho$. The asymmetry energy parameter $\beta_0$ yields a
relation between the constants $a^{\rm v}_{-}, h^{\rm v}_{1-}$ and $
h^{\rm v}_{2-}$. They are given in the upper half of Table 1. The
radius $r_0$ introduced above is shown instead of $\rho_0$. The
value used in Ref. \cite{Fay} was recalibrated in Ref.
\cite{Tol-Sap} to obtain a  more accurate description of nuclear
charge radii \cite{Sap-Tol}.
 One more remark should be made concerning the table. The ``natural'' TFFS normalizing factor
$C_0=2\eps_{0\rm F}/(3\rho_0)=308.2\;$MeV fm$^3$ corresponding to parameters of nuclear
matter in the third column of the table differs  from the one, $C_0=300\;$MeV fm$^3$, recommended
in the second edition of the Migdal's textbook on the TFFS \cite{AB2}. To make a comparison
with other articles within the TFFS, we recalculated all the strength parameters
to the latter. It explains a small difference of some values in the second column
in the table from the original those in  \cite{Fay}. An essential difference
between DF3 and DF3-a functionals takes place for the ``spin-dependent'' sector
in the bottom of the table. As we found in \cite{Tol-Sap}, the second one describes
the spin-orbit splitting of doublets better.

The anomalous component of the EDF \cite{Fay}reads
 \beq
 {\cal E}_{\rm an}({\bf r})=\sum_{\tau}{\cal
F}^{\xi,\tau\tau}({\bf r};[\rho])|\nu^{\tau}({\bf r})|^2,
\label{Ean}
\eeq
where the effective pairing interaction reads:
 \beq
{\cal F}^{\xi}=C_0 f^{\xi}=C_0 \left( f^{\xi}_{\rm ex}
+h^{\xi}x^{2/3}+f^{\xi}_{\nabla}r_0^2 (\nabla
x)^2\right).\label{Fxi}
 \eeq
 The first two terms are similar to
those in the TFFS \cite{Sap-Tr,Zv-Sap2} or in the SHFB method
\cite{Gor1}. The third in (\ref{Fxi}) is a new one introduced in
\cite{Fay5}. In this paper we use an approximate version of
(\ref{Fxi}) with $f^{\xi}_{\nabla}=0$. We will compare two models
for nuclear pairing: the ``volume'' pairing ($h^{\xi}=0$) and the
``surface'' pairing where both the pairing parameters $f^{\xi}_{\rm
ex}$ and $h^{\xi}$ are nonzero. One more remark should be made
concerning the pairing problem. In the approach \cite{Fay} pairing
was considered in the coordinate representation explicitly,
 solving the Gorkov equations with  the method developed in Ref.  \cite{Bel}.
 However, it was found that the results are practically
equivalent to those obtained within a  more simple BCS-like scheme
using the gap
 $\Delta_{\lambda \lambda'}=\Delta_{\lambda}\delta_{\lambda \lambda'}$
in a sufficiently large model space, $\eps_{\lambda}<E_{\rm max}$.
The effective pairing interaction (\ref{Fxi}) for the BCS approximation
is a little stronger than that in the coordinate representation (at
$\simeq 5\div 10$\%, depending on $E_{\rm max}$).
 For the systematic calculations in this article we use this simplified method of
considering the pairing problem with $E_{\rm max}=36\;$MeV. We do
not apply this method for nuclei close to the dripline for which the
diagonal approximation doesn't work \cite{Fay}.

%Table 1
\begin{table}[h!]
\caption{Parameters of the normal part of the EDF}
%\bigskip
\begin{tabular}{c|c|c}
\hline\hline
Parameter & DF3 \cite{Fay}& DF3-a \cite{Tol-Sap} \\
\hline

$\mu_0,\;$MeV& -16.05 & -16.05\\
$r_0,\;$fm& 1.147 & 1.145\\
$K_0,\;$MeV& 200 & 200 \\
$\beta,\;$MeV& 28.7 & 28.7 \\
\hline

$a^{\rm v}_+$    &-6.598    & -6.575   \\
$h^{\rm v}_{1+}$ & 0.163    &  0.163  \\
$h^{\rm v}_{2+}$ & 0.724    &  0.725  \\

$a^{\rm v}_-$    & 5.565    &  5.523  \\
$h^{\rm v}_{1-}$ &  0       &   0     \\
$h^{\rm v}_{2-}$ & 3.0      &  3.0    \\

$a^{\rm s}_+$    &-11.4     &  -11.1   \\
$h^{\rm s}_+$    & 0.31     &   0.31   \\

$a^{\rm s}_-$    &-4.11     & -4.10   \\
$h^{\rm s}_-$    &  0       &     0   \\
$r_{\rm c},\;$fm           & 0.35  &  0.35  \\
\hline

$\kappa$      &  0.216     & 0.190   \\
$\kappa'$     &  0.077     & 0.077   \\
$g_1$         &  0         & 0        \\
$g_1'$        & -0.123     & -0.308  \\
\hline
\hline
\end{tabular}
\end{table}

Within the TFFS, the response of a nucleus to the external quadrupole
field $V_0\exp{(i\omega t)}$ can be found in terms of the effective field.
In systems with pairing correlations, equation  for the effective
field can be written in a compact form as \beq {\hat
V}(\omega)={\hat V}_0(\omega)+{\hat {\cal F}}  {\hat A}(\omega) {\hat V}(\omega), \label{Vef_s}
\eeq where all the terms  are  matrices. In the
standard TFFS notation \cite{AB}, we have:
\beq {\hat V}=\left(\begin{array}{c}V
\\d_1\\d_2\end{array}\right)\,,\quad{\hat
V}_0=\left(\begin{array}{c}V_0
\\0\\0\end{array}\right)\,,
\label{Vs} \eeq

\beq {\hat {\cal F}}=\left(\begin{array}{ccc}
{\cal F} &{\cal F}^{\omega \xi}&{\cal F}^{\omega \xi}\\
{\cal F}^{\xi \omega }&{\cal F}^\xi  &{\cal F}^{\xi \omega }\\
{\cal F}^{\xi \omega }&{\cal F}^{\xi \omega }& {\cal F}^\xi \end{array}\right), \label{Fs} \eeq

\beq {\hat A}(\omega)=\left(\begin{array}{ccc} {\cal L}(\omega) &{\cal M}_1(\omega)
&{\cal M}_2(\omega)\\
 {\cal O}(\omega)&-{\cal N}_1(\omega) &{\cal N}_2(\omega)\\{\cal O}(-\omega)&-{\cal N}_1(-\omega) &
 {\cal N}_2(-\omega)
\end{array}\right)\,,
\label{As} \eeq where ${\cal L},\; {\cal M}_1$, and so on stand
for integrals over $\eps$ of the products of different
combinations of the Green function $G(\eps)$ and two Gor'kov
functios $F^{(1)}(\eps)$ and $F^{(2)}(\eps)$. They can be found in
\cite{AB} and we write down here only the first of them which is
of the main importance for us, \beq {\cal
L}=\int\frac{d\varepsilon}{2\pi
i}\left[G(\varepsilon)G(\varepsilon+\omega)-F^{(1)}(\varepsilon)F^{(2)}(\varepsilon+\omega)
\right]. \label{Ls} \eeq

Isotopic indices in Eqs. (\ref{Vs}-\ref{As}) are omitted for
brevity. In Eq. (\ref{Fs}), ${\cal F}$ is the usual LM amplitude,
\beq {\cal F}=\frac {\delta^2 {\cal E}}{\delta \rho^2}, \label{LM}
\eeq whereas the amplitudes ${\cal F}^{\omega \xi}={\cal F}^{\xi
\omega}$ stand for the mixed second derivatives, \beq {\cal
F}^{\omega \xi}=\frac {\delta^2 {\cal E}}{\delta \rho \delta \nu}.
\label{LMxi} \eeq  In the case of volume pairing, we have ${\cal
F}^{\omega \xi}=0$. The explicit form of the above equations and
(\ref{Ls}) is written down for the case of the electric ($t$-even)
symmetry we deal with. A static moment of an odd nucleus can be
found in terms of the diagonal matrix element $ \langle\lambda_0|
V(\omega =0)|\lambda_0\rangle$ of the effective field over the state
$\lambda_0$ of the odd nucleon.

The effective field operator ${\hat V}(\omega)$ has a pole in the excitation
energy $\omega_s$ of the state $|s\rangle$ under consideration,
\begin{equation}\label{pole}
    {\hat V}(\omega)=\frac {\left({\hat V_0}{\hat A(\omega_s)} {\hat g}_{0s}
     \right){\hat g}_{0s}}{\omega-\omega_s}+{\hat V}_R(\omega).
\end{equation}
The quantity ${\hat g}_{0s}$ has the meaning of the corresponding
excitation amplitude. It obeys  the homogeneous counterpart of Eq.
(\ref{Vef_s}) and is normalized as follows \cite{AB},
\begin{equation}\label{norm}
\left({\hat g}_{0s}^+ \frac {d {\hat A}}{d\omega}{\hat g}_{0s} \right)_{\omega=\omega_s}=-1,
\end{equation}
with obvious notation.

For excitation probabilities, it is more convenient to use the transition density operator
which is conjugated to ${\hat g}_{0s}$,
\begin{equation}\label{rhot}
    {\hat \rho}^{\rm tr}_{0s} = {\hat A}{\hat g}_{0s}.
\end{equation}

The explicit definition of the normal and anomalous components of
${\hat \rho}^{\rm tr}_{0s}$ is as follows
\begin{equation}\label{rho0}
\rho^{{\rm tr}(0)}_{0s}({\bf r},{\bf r}')=\int \frac {d\eps}{2\pi i}
\delta G({\bf r},{\bf r}';\eps,\omega_s),
\end{equation}
\begin{equation}\label{rho12}
\rho^{{\rm tr}(1,2)}_{0s}({\bf r},{\bf r}')=\int \frac {d\eps}{2\pi
i} \delta F^{(1,2)}({\bf r},{\bf r}';\eps,\omega_s).
\end{equation}

The TFFS equation for transition densities for nuclei with pairing correlations,
\begin{equation}\label{rhotr}
{\hat \rho}^{\rm tr}_{0s} = {\hat A}(\omega_s) {\hat {\cal F}} {\hat
\rho}^{\rm tr}_{0s},
\end{equation}
is a complete analogue of the QRPA set of equations. Therefore we will often name it
the QRPA equation.
The transition density is normalized due to Eq. (\ref{norm}), and the transition matrix element
for the excitation of the state $|s\rangle$ with
the external field $V_0$ is given by
\begin{equation}\label{M0s}
M_{0s}=\int {\hat V}_0 {\hat \rho}^{\rm tr}_{0s}({\bf r}) d{\bf r}.
\end{equation}

\section{Characteristics of the $2^+_1$ excitations}
The formalism described in the previous Section was used to describe
$2^+_1$ states in two isotopic chains of semi-magic nuclei, lead and
tin. We investigate both a   pure surface and a pure volume version
of pairing.  More calculational details can be found in Ref.
\cite{Fay}. We use the so-called developed pairing approximation. In
particular, we don't make any corrections to particle
non-conservation effects induced with the Bogolyubov transformation.
Therefore  in the vicinity of double magic nuclei, the results
should be considered as very approximate. As it was found in
\cite{Fay}, it is impossible to describe neutron and proton
separation energies $S_n$ and $S_p$ for all nuclei, from calcium up
to  lead, with sufficient accuracy using a  fixed set  of parameters
in Eq. (\ref{Fxi}), the effective strength of the pairing
interaction should be diminished with increasing nucleon number A.

\begin{figure}[ht!]
\centerline {\includegraphics [width=80mm]{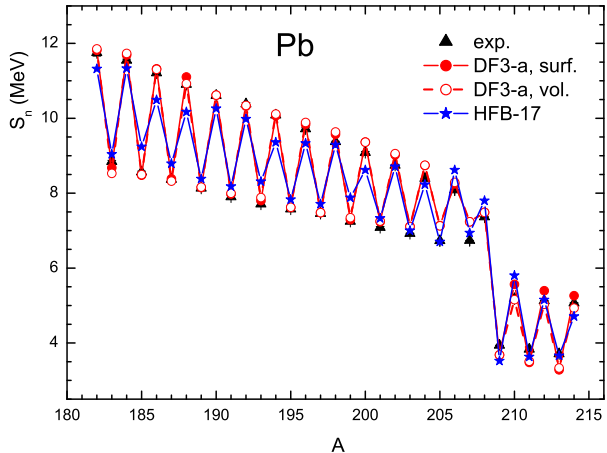}} \vspace{2mm}
\caption{(Color online) Neutron separation energies $S_n$ for lead isotopes. The
volume pairing corresponds to ($f^{\xi}_{\rm ex}=-0.31;h^{\xi}=0$),
the surface one, to ($f^{\xi}_{\rm ex}=-1.05;\;h^{\xi}=0.94$). The
HFB theory predictions with the HFB-17 Skyrme functional are taken
from \cite{site}.}
\end{figure}

\begin{figure}[ht!]
\centerline {\includegraphics [width=80mm]{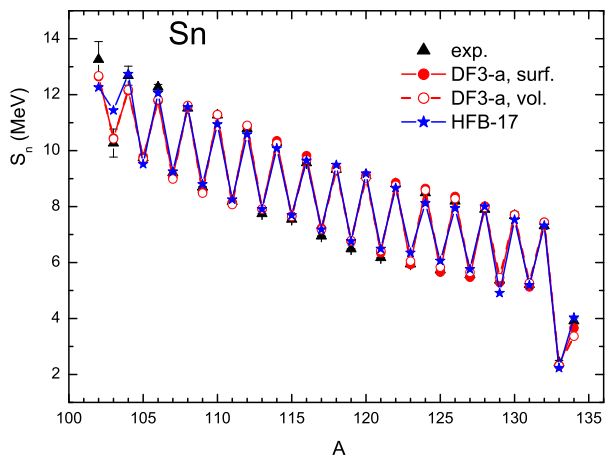}} \vspace{2mm}
\caption{(Color online) Neutron separation energies $S_n$ for tin
isotopes. The volume pairing corresponds to ($f^{\xi}_{\rm
ex}=-0.33;\;h^{\xi}=0$), the surface one, to ($f^{\xi}_{\rm
ex}=-1.05;\;h^{\xi}=0.92$). The HFB theory predictions with the
HFB-17 Skyrme functional are taken from \cite{site}.}
\end{figure}

In this paper, we limit ourselves to  two long isotopic chains, the
lead and the tin chains. Therefore we deal with neutron pairing
only. A short comment should be made on the procedure of solving the
pairing problem.  No particle number projection procedure is used in
our calculations, i.e. particle number is conserved only on average,
corresponding to the chosen chemical potential $\mu$ for the kind of
nucleons under consideration. The accuracy of this approximation is
examined in a lot of papers. For the self-consistent SHF method with
the SLy4 force, it was found in recent article \cite{part-numb} that
the average difference between exact and approximate gap values is
0.12 MeV, the error being bigger in vicinity of magic nuclei.

For finding the parameters of the pairing force (\ref{Fxi}) we use
the strategy of ``soft'' variation of them  to obtain better values
of $S_n$ for both the chains under consideration. Values of $S_n$
for both kinds of pairing are compared with the data in Fig. 1 and
Fig. 2. Explicit values of the pairing parameters are given in the
figure captions. Remind that we use the two-parameter version of
(\ref{Fxi}), with $f^{\xi}_{\nabla}=0$. For the volume pairing
($h^{\xi}=0$), one parameter remains which is smaller for lead than
for tin approximately at 6\%. For the surface pairing we deal with a
two-parameter form of ${\cal F}^{\xi}$. The ``external'' pairing
parameter $f^{\xi}_{\rm ex}$ is taken $A$-independent, in accordance
with its physical meaning as the free NN $T$-matrix taken at
negative energy $E=2\mu$ \cite{Bald1}. As to the second one,
$h^{\xi}$, it increases  from the Sn chain to the Pb one at 2\%, the
resulting pairing attraction again becoming weaker, but only a
little. Thus, the $A$-independence of the pairing parameters in the
case of surface pairing is weaker than for the volume one. This
finding suggests to favor  surface pairing.
 As we see, the difference between the predictions
for neutron separation energies is small for
both versions and agreement with the experimental data is nearly perfect.
For comparison, we display the predictions of the HFB-17 version
of the Skyrme force \cite{Gor1} which  has a record accuracy in
overall description of nuclear masses. We see that for these two
chains our accuracy in description of neutron separation energies is
even better. Of course, we achieved the agreement by a small
variation of one of two paring parameters whereas calculations
\cite{Gor1} are carried out with an universal set of parameters.
However, the pairing part of the HFB-17 functional contains five
parameters.

\begin{figure}[ht!]
\centerline {\includegraphics [width=80mm]{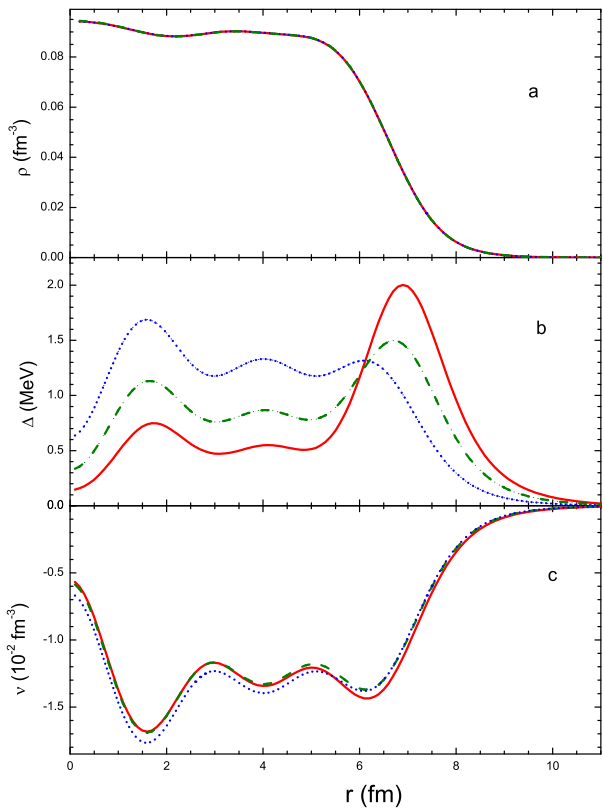}} \vspace{2mm}
\caption{(Color online) Neutron density (pannel a), gap (b) and anomalous density
(c) in $^{200}$Pb nucleus. Solid and dotted lines correspond to the
surface and volume versions correspondingly, the dashed one, to the
medium version ($f^{\xi}_{\rm ex}=-0.70;\;h^{\xi}=0.50$).}
\end{figure}

Fig. 3 demonstrates that the normal neutron density $\rho_n(r)$ and the anomalous one,
$\nu_n(r)$, both are practically insensitive to the kind of paring used in the calculation.
On the contrary, the gap itself is very sensitive.
For comparison, we took also a ``medium'' version, with ($f^{\xi}_{\rm ex}=-0.70;\;h^{\xi}=0.50$).
It gives $S_n$ value approximately with the same accuracy as the previous two.

\begin{figure*}
\centerline {\includegraphics [width=140mm]{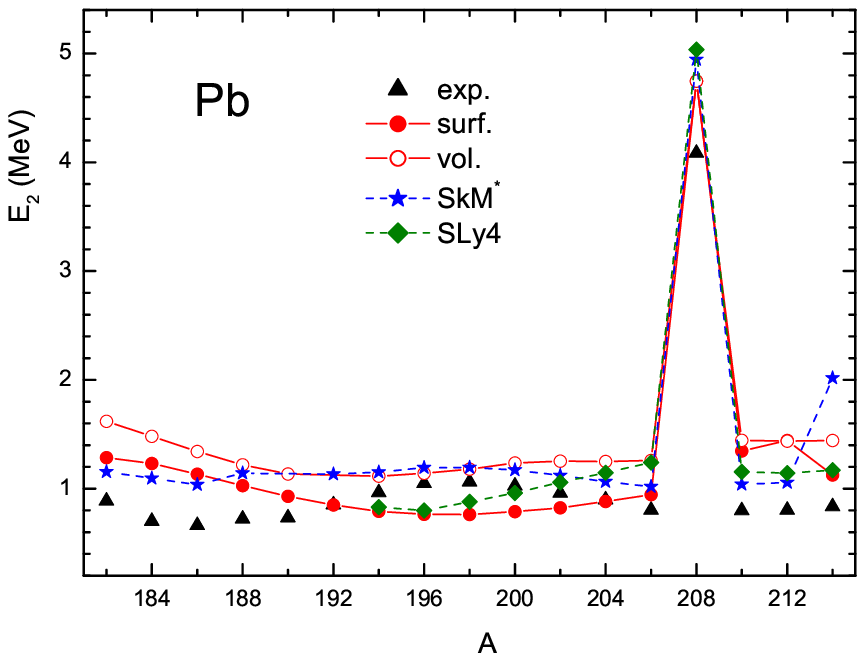}} \vspace{2mm}
\caption{(Color online) Excitation energies $\omega(2^+_1)$ for lead isotopes.
Predictions for mean field approach with the forces SkM*(dashed blue
line) and  SLy4(dashed green line)  are taken from \cite{Bertsch1}.
The energy density functional results are given by the solid lines.}
\end{figure*}

\begin{figure*}
\centerline {\includegraphics [width=140mm]{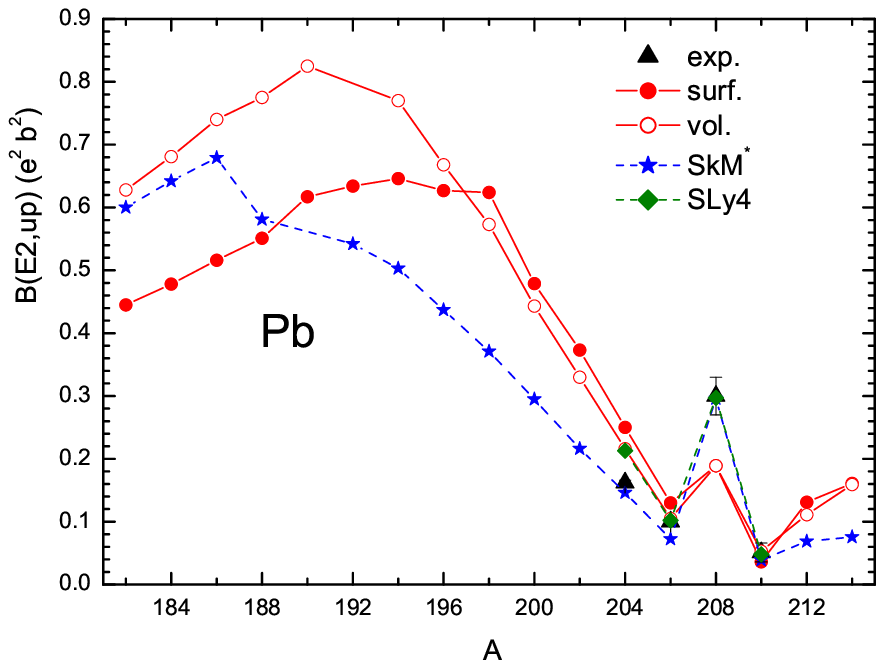}} \vspace{2mm}
\caption{(Color online)$B(E2,{\rm up})$ values for lead isotopes.  Predictions
for the SkM* and SLy4 force are taken from \cite{Bertsch1}.}
\end{figure*}

Let us now examine
to what extent predictions for characteristics of $2^+_1$ states  are different for these two
versions of pairing force which are equivalent
in describing the $S_n$ values. A comment should be made before
presenting results of the QRPA calculations.
Our QRPA code doesn't include the spin-orbit (\ref{Fsl}) and
spin (\ref{g1}) terms of the effective interaction, therefore the self-consistency
 is not complete and the excitation energy of the ghost $1^-$-state does  not  automatically vanish.
In the present investigation, we fine tuned
the parameter $a^{\rm v}_+$ in (\ref{EDF_v})
in order to decouple the ghost state. This change is different for
different nuclei but on average the value of $|a^{\rm v}_+|$ increases at $\simeq 3$\%
in comparison with that given in Table 1.

Let us begin with the lead chain. Excitation energies $\omega_{2}$
are displayed in Fig. 4 and the probabilities $B(E2,{\rm up})$, in
Fig. 5. Experimental data for both quantities are taken from
\cite{Dat}. For comparison, results of the QRPA calculations of
\cite{Bertsch1} with the SkM* and SLy4 force are shown. Note that
they were carried out with  density independent pairing. We see that
the difference $\delta \omega_{2}=\omega_{2}^{\rm vol} -
\omega_{2}^{\rm surf}$ is, on average, of the order of 0.3 MeV, that
is the effect under discussion  is noticeable for this quantity. The
results for volume pairing  are systematically higher, with the
exception of the $^{210,212}$Pb isotopes for which the two versions
practically coincide. Agreement with the data is, on average, quite
reasonable for both the versions. Predictions of both the SkM* and
SLy4 QRPA calculations for $\omega_{2}$ values have approximately
the same accuracy as ours.

For excitation probabilities the situation is more complex.
For isotopes heavier than $^{198}$Pb our ``surface''  and ``volume''
curves are very close to each other. For lighter part of the chain
the volume pairing generates larger probabilities than surface pairing does,
producing differences up to  $\simeq 30$\%.
 Comparing with Fig. 4, we see that  there is some
unusual correlation between  excitations energies and probabilities.
Indeed, in magic nuclei where the pairing is absent for low-lying
collective excitations there   is a rule that a lower  energy
implies a larger probability. It can be qualitatively
explained with the hydrodynamical Bohr-Mottelson (BM) model
\cite{BM2} which gives a simple relation
for the transition density of a $L$-vibration:
\begin{equation}\label{rhoBM}
    \rho_L^{\rm tr,BM}=\alpha_L \frac {d\rho}{dr},
\end{equation}
where $\alpha_L=1/\sqrt{2\omega_L B_L}$, and $B_L$ is the collective mass parameter of the BM model
 proportional to the nuclear mass. Then one obtains
\beq \label{BELBM}
 B(EL,{\rm up})=\frac{2L+1}{2\omega_L B_L} (M_L)^2,
\eeq where $M_L^{\rm BM}=(3Ze/4\pi)R^{L-1}$, $R$ being the nuclear
radius. Thus, in the BM model a lower value of the excitation energy
$\omega_L$ inevitably leads to a higher value of the excitation
probability. In our calculations, the situation is opposite. In
principle, this is not strange. Indeed, even in  magic nuclei the BM
model works  only qualitatively \cite{KhS}. If one solves equations
of the self-consistent TFFS or any HF+RPA equations for nuclei
without pairing, Eq. (\ref{rhoBM}) remains approximately true, but
the mass parameter becomes $\omega$-dependent and deviates from the
BM model prescription significantly \cite{KhS}. In nuclei with
pairing, the situation becomes even more different from this
simplest model as the normal component of the transition density
(\ref{rho0}) depends now from the anomalous transition amplitudes
${\hat g}_{0s}^{(1,2)}$ (see Eq. (\ref{rhotr})). They strongly
depend on the kind of pairing. As a result, the correlation
between the $\omega_L$ and $B(EL)$ values of the BM-type (\ref{BELBM})
can be destroyed.

Experimental probabilities are known only for four even
$^{204-210}$Pb isotopes. For all of them, the SkM* and SLy4
calculations are in perfect agreement with the data. Agreement of
our calculations is poorer. It is especially true for the magic
$^{208}$Pb nucleus where there is no any pairing. It should be noted
that in this nucleus the collectivity of the $2^+_1$-state is not
high: the B(E2) value is only about 8 single-particle units (spu).
For a comparison, the B(E3) value for the $3^-_1$-state exceeds 30
spu. But for excitations with low collectivity in nuclei without
pairing the RPA solution depends strongly on the single-particle
spectrum, and even a small inaccuracy in the positions of
single-particle levels can change results significantly.  In any
case, some modification of the normal part of the functional DF3-a
is necessary to obtain better agreement for the $^{208}$Pb nucleus.

\begin{figure*}
\centerline {\includegraphics [width=140mm]{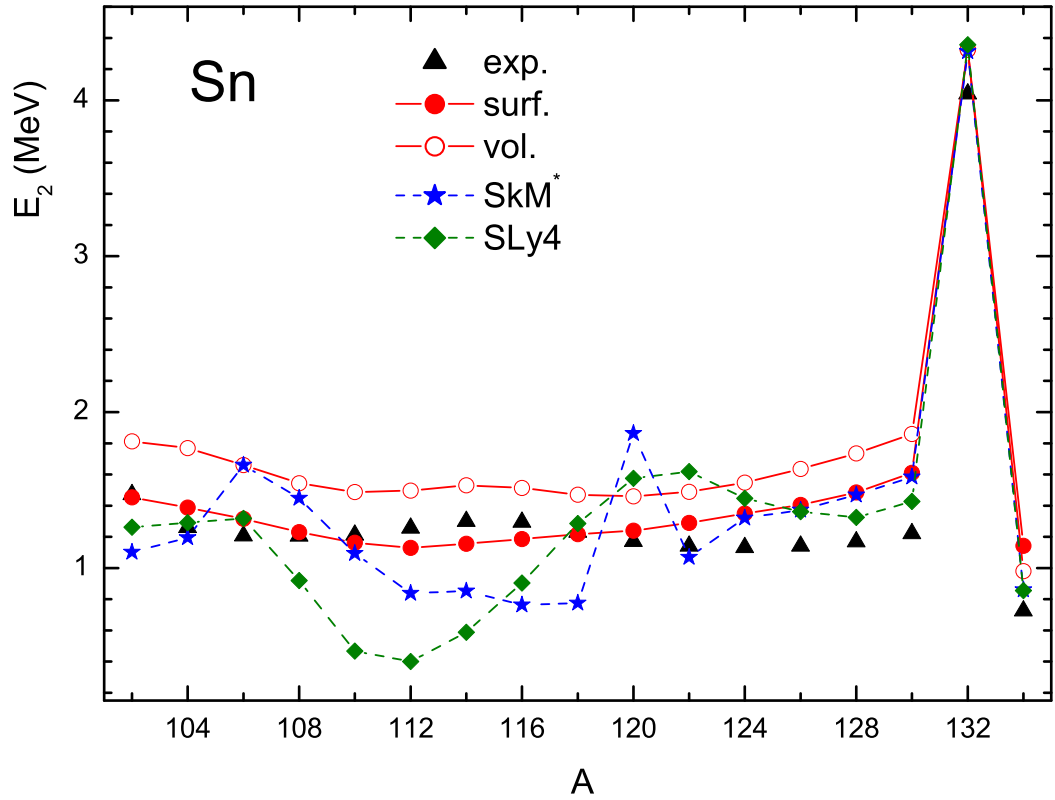}} \vspace{2mm}
\caption{(Color online) Excitation energies $\omega(2^+_1)$ for tin isotopes.
Predictions for the SkM* and SLy4 force are taken from \cite{Bertsch1}.}
\end{figure*}

In the tin chain, see Fig. 6, the situation with the excitation
spectrum is partially similar to the one in lead. Again,
$2^+_1$-levels  are higher for volume pairing than in the surface
case, and again the difference $\delta \omega_{2}$ is $\simeq
300\;$keV, the surface predictions being closer to the experimental
data. As to the SkM* spectrum, for isotopes heavier than $^{122}$Sn
it practically coincides with our ``surface'' one, both being higher
than the experimental spectrum by approximately $200\div 300\;$keV.
For lighter isotopes, it deviates from our surface spectrum
significantly in an irregular way whereas the latter practically
coincides with the experiment in this $A$ region. As to the SLy4
spectrum, it also looks reasonable for the heavy part of the chain
but for isotopes lighter of $^{124}$Sn it strongly oscillates around
the experimental curve. In the dip minimum for $^{112}$Sn the
$\omega_{2}$ value is less than the experimental one at
approximately 1 MeV and it is close to an instability.

The excitation probabilities are displayed in Fig. 7. Here the
results show a  very complex pattern. For the heavier
part of the chain, beginning at the $^{124}$Sn nucleus,  our two
theoretical curves and the SkM* practically coincide, all being
close to the experiment.
 The SkM* curve
behaves in a non-regular way with strong deviations from the
experimental data, up to $\simeq 50\div 100$\%. The Sly4
interaction produces  excitation probabilities which strongly
decrease with the nucleon number A, implying drastic deviations
from the data. The density functional approach is able to describe
the A-dependence of the experimental $B(E2,{\rm up})$ values
rather well.
 For lighter tin isotopes, our two curves
began to deviate from each other, the volume one being higher by
$\simeq 25\div 30$\%, and  a first glance may suggest
 that the volume pairing interaction performs much better.
On the other hand, one has to notice the large error bars of the
experimental data in the mass region below A=114.

\begin{figure*}
\centerline {\includegraphics [width=140mm]{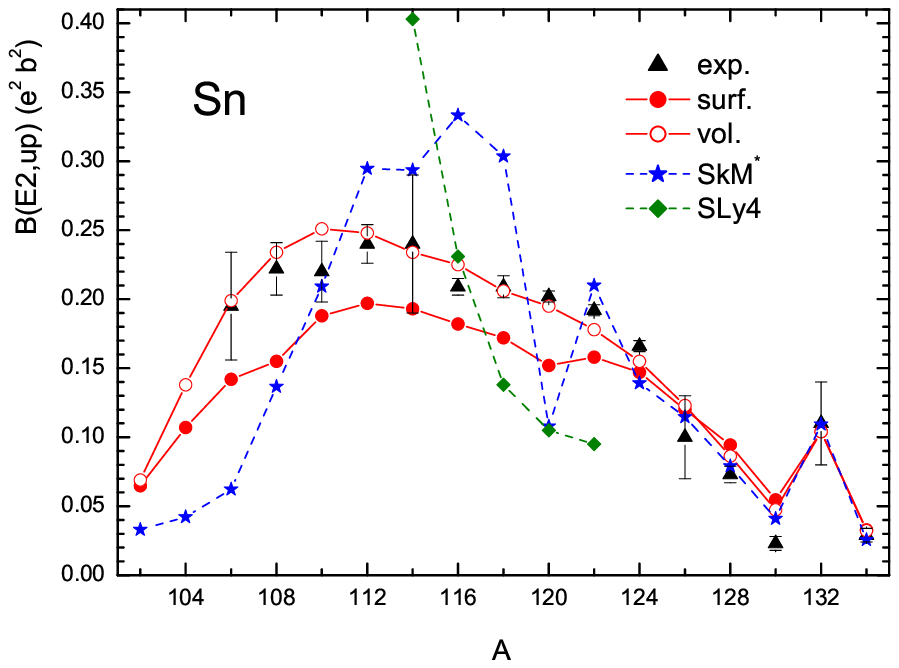}} \vspace{2mm}
\caption{(Color online) $B(E2,{\rm up})$ values for tin isotopes.  Predictions
for the SkM* and SLy4 force are taken from \cite{Bertsch1}.
Experimental data are taken for $^{114-124}$Sn from \cite{Dat},
for $^{126-134}$Sn  from \cite{newbe2}, and for $^{106-112}$Sn
from \cite{be2_106-112,be2_110,be2_106-108}.}
\end{figure*}

To investigate the role of pairing itself and of the type of its
density dependence in detail, let us analyze different components of
the transition amplitude. Let us begin from the anomalous terms
${\hat g}^{(1,2)}$ (the index ``0s'' is for brevity omitted). They
are displayed in Fig. 8 for the $^{200}$Pb nucleus. We see, first,
that, for both the versions, the $g^{(1)}$ amplitude value is much
bigger than $|g^{(2)}|$.  Second, the coordinate dependence of the main
$g^{(1)}$ amplitude is absolutely different for the two versions
under comparison. In the surface pairing case,  a strong surface
maximum dominates whereas in the volume case $g^{(1)}$  is spread
over the volume, with rather strong oscillations. In addition, it is
seen that the integral effect of $g^{(1)}_{\rm surf}$ should be
noticeably bigger than that of $g^{(1)}_{\rm vol}$.  All this shows
some asymmetry for Bogolyubov quasiparticles and quasiholes. Such a
situation is typical for nuclei which are close to the magic core.

\begin{figure}[ht!]
\centerline {\includegraphics [width=80mm]{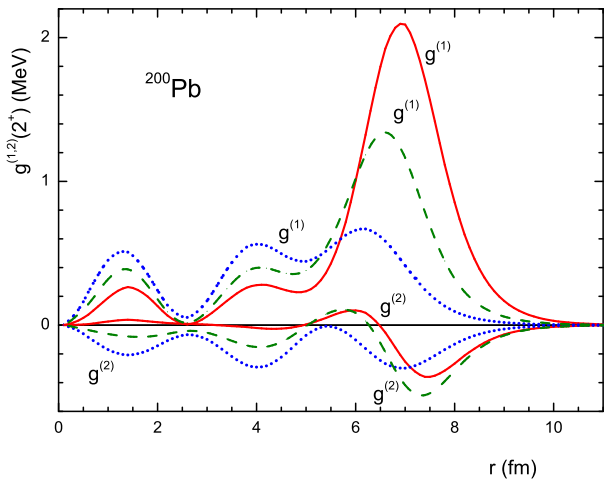}} \vspace{2mm}
\caption{(Color online) The neutron anomalous transition amplitudes $g^{(1,2)}$ in
$^{200}$Pb nucleus. Solid lines correspond to
 surface, dotted  to volume, and dashed, to the medium kind of pairing, see Fig. 3.}
\end{figure}

The normal proton and neutron amplitudes $g^{(0)}$ for the same
nucleus are displayed in Fig. 9. As we see, for this quantity
the influence of the kind of pairing used is minimal. Thus, evidently,
the rather big value  of the difference $\delta \omega_{2}\simeq 300$
keV for this nucleus is explained with different contributions of
the anomalous amplitude $g^{(1)}$ which is much stronger in the case
of surface pairing. For the transition densities, see Fig. 10, the
effect is rather small but a little bigger than for the normal
amplitudes $g^{(0)}$. This additional enhancement of the surface
maximum of $\rho^{tr(0)}(r)$  in the surface pairing case again
originates from the term with $g^{(1)}$ in Eq. (\ref{rhot}). In its
turn, it explains the increase of the  $B(E2)$ value in this
nucleus for the surface case.

\begin{figure}[ht!]
\centerline {\includegraphics [width=80mm]{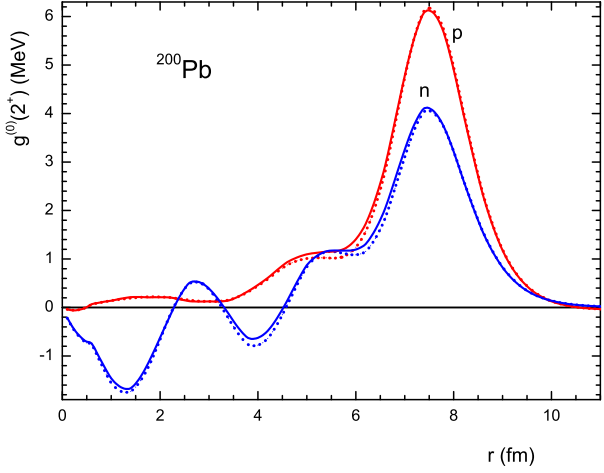}} \vspace{2mm}
\caption{(Color online) The proton and neutron normal transition amplitudes
$g^{(0)}$ in $^{200}$Pb nucleus. Solid lines correspond to surface
pairing, dotted ones, to volume pairing.}
\end{figure}

\begin{figure}[ht!]
\centerline {\includegraphics [width=80mm]{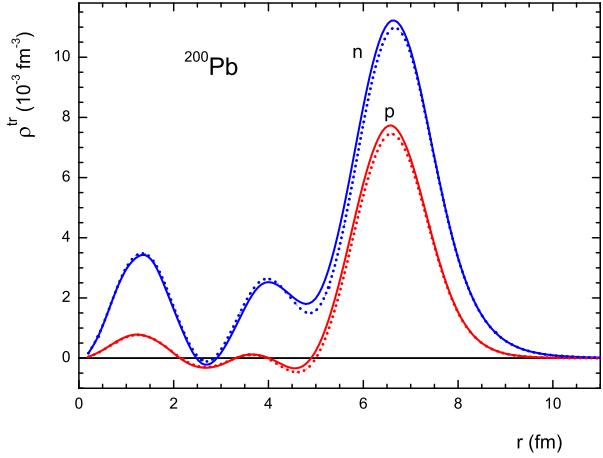}} \vspace{2mm}
\caption{(Color online) The proton and neutron  transition densities
$\rho^{tr(0)}$ in $^{200}$Pb nucleus. Solid lines correspond to
surface pairing, dotted ones, to volume pairing.}
\end{figure}

Let us go to the tin chain.  Figs. 11--13 present for the $^{118}$Sn
nucleus the same quantities which were displayed in Figs. 8--10 for
the $^{200}$Pb nucleus. This nucleus  is in the middle of the chain,
and all properties of the ``developed'' pairing, in particular,
particle-hole symmetry should take place. Indeed, now (see Fig. 11)
the amplitudes $g^{(1)}$ and $g^{(2)}$ possess a similar form and
absolute value and, being of the opposite sign. In the result, we
have $|g^{(-)}=g^{(1)}-g^{(2)}| \gg |g^{(+)}=g^{(1)}+ g^{(2)}|$ as
it should be \cite{AB}. Again, as in the $^{200}$Pb case, the effect
of the kind of pairing on the magnitude of  $g^{(1,2)}$ is drastic.
As to that for the normal amplitudes $g^{(0)}$ and transition
densities $\rho^{tr(0)}$, again it is rather moderate but of the
another sign. Now in the volume case, the surface peaks in both
these quantities are higher and, correspondingly, the $B(E2)$ value
is bigger. Evidently, in this case we deal with some destructive
interference between normal and anomalous contributions to solutions
of the equations of Section 2.

To summarize, we see an  effect of the type of pairing on
the characteristics of the  $2^+_1$-states in spherical nuclei. The
excitation energies $\omega_2$ are systematically lower in the
surface case to $\delta \omega_{2}\simeq 300\;$keV, and the surface
values are, as a rule, closer to the data. For $B(E2)$ values, the
effect is not so regular and here the volume version predictions on
average look better.
Thus, the present analysis is compatible with both volume and surface
pairing.

\begin{figure}[ht!]
\centerline {\includegraphics [width=80mm]{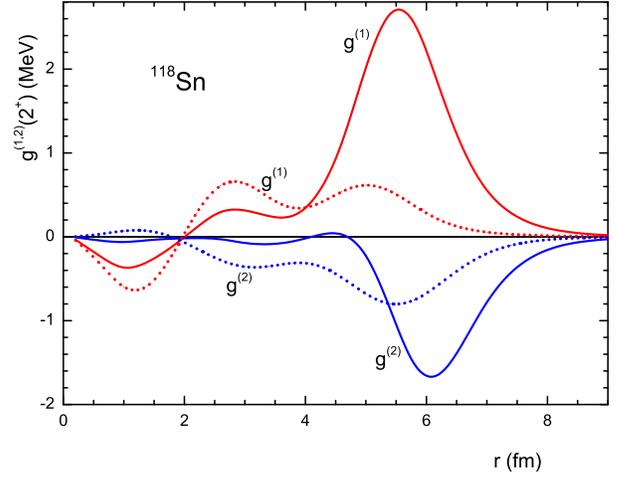}}
\vspace{2mm} \caption{(Color online) The neutron anomalous transition amplitudes
$g^{(1,2)}$ in $^{118}$Sn nucleus. Solid lines correspond to
 surface pairing, dotted ones, to volume pairing.}
\end{figure}

\begin{figure}[ht!]
\centerline {\includegraphics [width=80mm]{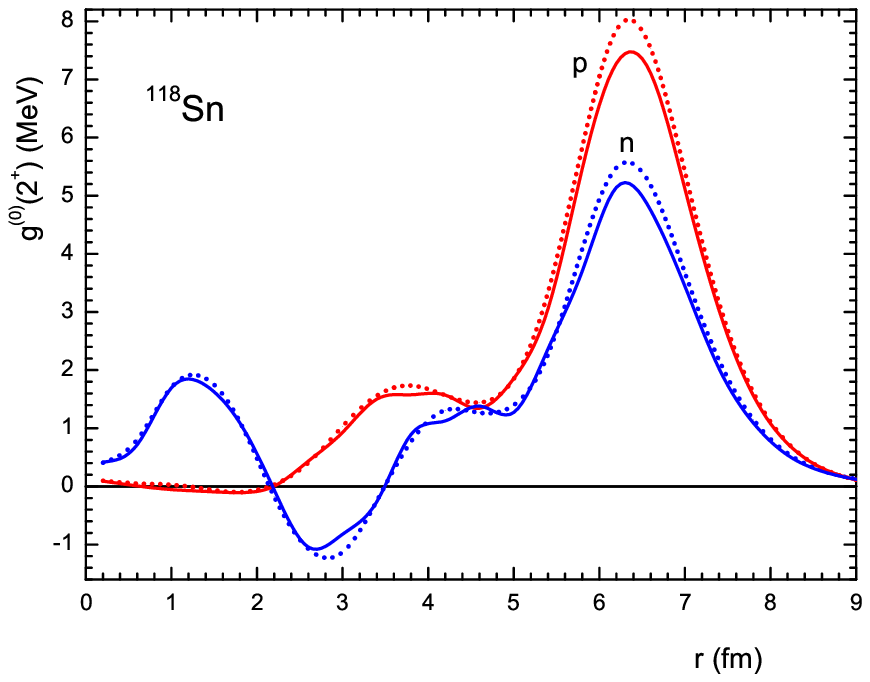}}
\vspace{2mm} \caption{(Color online) The same as in Fig. 9 but for the $^{118}$Sn
nucleus.}
\end{figure}

\begin{figure}[ht!]
\centerline {\includegraphics [width=80mm]{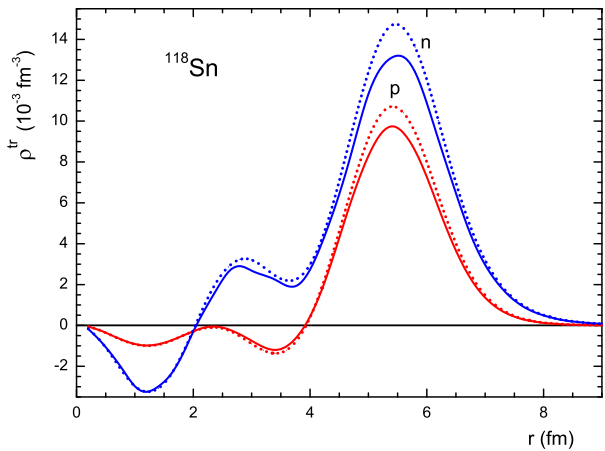}} \vspace{2mm}
\caption{(Color online) The same as in Fig. 10 but for the $^{118}$Sn nucleus.}
\end{figure}

The issue could be naturally raised to what extent the small differences
seen in the observables can be traced to the kind of pairing employed.
 In other words, is it possible to fine tune the interaction parameters
such that the both volume and surface pairing produce
indistiguishable results, while keeping the mass differences and
separation energies close to the experimental data? We carried out
such an analysis for the tin chain. We consider the double mass
differences \beq D(N)=\frac 1 2 \left(S_n(N)-\frac 1  2
\left(S_n(N-1)+S_n(N+1)\right)\right), \eeq $N$ even, which is very
sensitive to the value of pairing gap. Note that the approximate
relation $D(N)\simeq  \bar{\Delta}$ takes place where $\bar{\Delta}$
is an average gap value. We calculate the average difference between
theoretical and experimental values of this quantity, \beq \langle
\delta D \rangle =  \sqrt{\frac 1 N \sum_{N} \left(D_{\rm th}(N)-
D_{\rm exp}(N)\right)^2}, \eeq $N$ even. We include into the
analysis isotopes from $^{106}$Sn till $^{128}$Sn for which the
developed pairing approximation seems to be reasonable. Results are
presented in Table II, for the surface pairing and for three
versions of the volume pairing with different values of the strength
parameter $f^{\xi}$. For all of them the characteristics of the
$2^+_1$-state in the example $^{118}$Sn nucleus are given. It is
seen that with increase of $|f^{\xi}|$ from the optimal value
$f^{\xi}=-0.33$ deviations from the surface version predictions
grow. With decrease of $|f^{\xi}|$ they become less, but this effect
is much less than the initial difference even for the value
$f^{\xi}=-0.32$ for which description of the mass differences is
essentially worse than for the optimal value. An additional decrease
of  $|f^{\xi}|$ will absolutely destroy the nuclear mass
description. In  other isotopes of the tin chain, influence of
variation of the $f^{\xi}$ parameter to  values of $\omega_{2}$ and
$B(E2)$ is quite similar. Thus, the effect under discussion
originates mainly due to the surface nature of pairing {\it versus}
the volume one.

\begin{table}[t]
\caption{Dependence of the $2^+_1$-state characteristics of the
$^{118}$Sn nucleus on the strength of pairing force.}

\begin{tabular}{c c c c   }
\hline \hline

version  & $\langle \delta D \rangle$ (MeV)  & $\omega_{2}$ (MeV)
& \hspace*{1.5ex} $B(E2,{\rm up)}$(${\rm e^2 b^2}$)\hspace*{1.5ex}  \\
\hline

surface              & 0.10 & 1.216 & 0.172 \\

vol. $f^{\xi}=-0.33$ & 0.11 & 1.470 & 0.206 \\

vol. $f^{\xi}=-0.32$ & 0.16 & 1.375 & 0.193 \\

vol. $f^{\xi}=-0.34$ & 0.11 & 1.570 & 0.216 \\

\hline \hline

\end{tabular}\label{tab_Qp}
\end{table}

In conclusion of this Section we compare in Fig. 14 the charge
transition density $\rho^{\rm tr}_{\rm ch}(r)$ in the $^{118}$Sn
nucleus with the experimental transition charge density found with
a model independent analysis of the elastic electron scattering in
\cite{rhotr}. The theoretical charge density is obtained  from
$\rho^{\rm tr}_p(r)$ and $\rho^{\rm tr}_n(r)$ functions displayed
in Fig. 13 with taking into account relativistic corrections
\cite{Friar}. For both  versions of pairing the agreement with
the data is quite reasonable, and it is a little better in the
surface case.

\begin{figure}[ht!]
\centerline {\includegraphics [width=80mm]{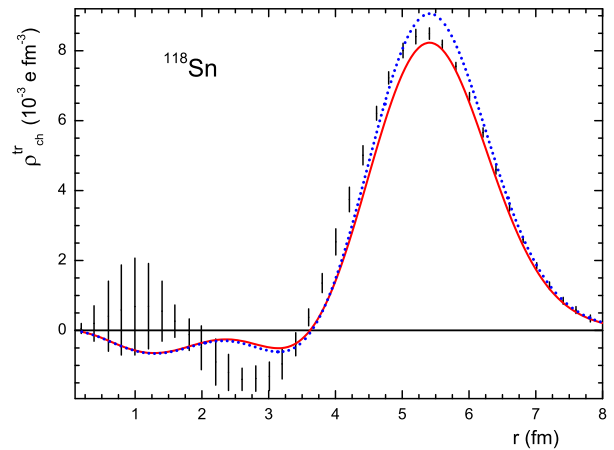}} \vspace{2mm}
\caption{(Color online) The charge transition densities $\rho_{\rm
ch}^{{\rm tr}(0)}$ in $^{118}$Sn  nucleus. Solid lines correspond to
surface pairing, dotted ones, to volume pairing.}
\end{figure}

\section{Quadrupole moments of odd nuclei}

\begin{figure}[ht!]
\centerline {\includegraphics [width=80mm]{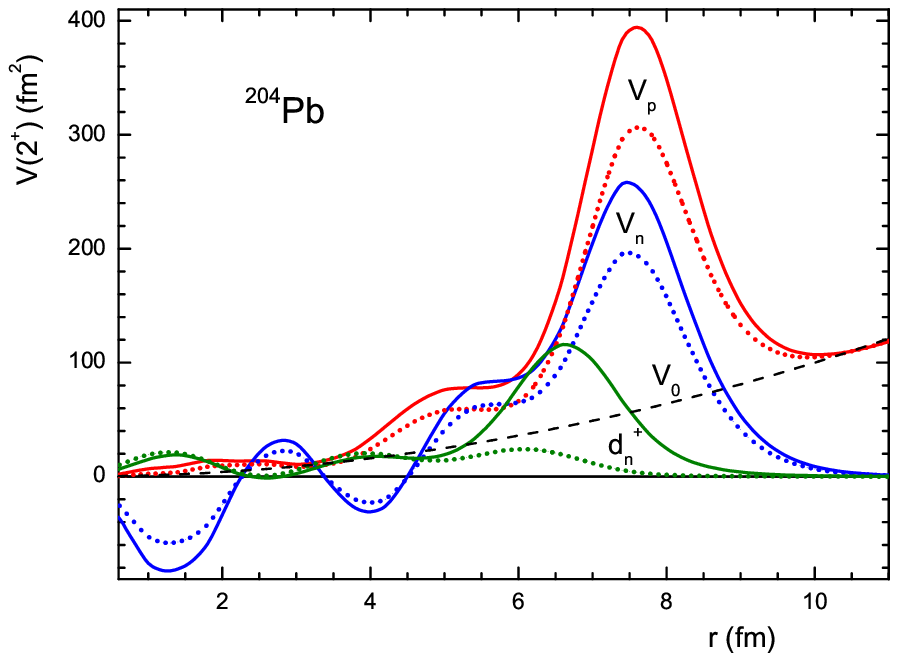}} \vspace{2mm}
\caption{(Color online) Static effective fields $V_n$ $V_p$ and $d^+_n$ in
$^{204}$Pb nucleus. Solid lines correspond to surface pairing,
dotted ones, to volume pairing.}
\end{figure}

\begin{figure}[ht!]
\centerline {\includegraphics [width=80mm]{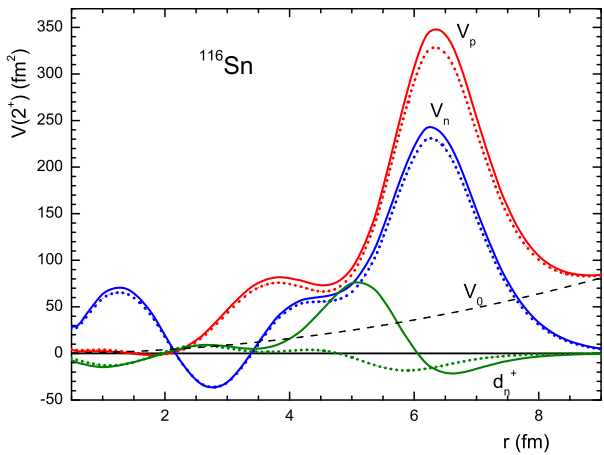}} \vspace{2mm}
\caption{(Color online) Static effective fields $V_n$ $V_p$ and $d^+_n$ in
$^{116}$Sn nucleus. Solid lines correspond to surface pairing,
dotted ones, to volume pairing.}
\end{figure}

 Recently magnetic moments of odd spherical nuclei
have been calculated  \cite{borzov2008,borzov2010} within the same
self-consistent approach as the one used here. A reasonable
description of the data for more than  hundred of spherical nuclei
was obtained. Especially high accuracy was reached for semi-magic
nuclei considered in the ``single-quasiparticle approximation''
where one quasiparticle in the fixed state $\lambda=(n,l,j,m)$ with
the energy $\varepsilon_{\lambda}$ is added to the even-even core.
 According to the TFFS,  a quasiparticle  differs  from
 a particle of the single-particle model in two respects.
First, it possesses the local charge $e_q$
 (in our case,  we have
$e_{q}^{p}$ = 1, $e_{q}^{n}$ = 0), and, second, the core is polarized
 due to the interaction between the particle and the core nucleons via the
 LM amplitude. In other words, the quasiparticle possesses the
 effective charge $e_{\rm eff}$ caused by the polarizability  of the core which is found
 by solving the above TFFS equations. In the many-particle Shell Model \cite{BABr},
 a similar quantity is  introduced as a phenomenological parameter which describes
 polarizability of the core consisting of  outside nucleons.

 In non-magic nuclei,  the term quasiparticle
takes a double meaning. In addition to the initial LM concept we
consider the Bogolyubov quasiparticles with occupation numbers
$n_{\lambda}^{\rm B}{=}(E_{\lambda}-\eps_{\lambda})/2E_{\lambda}$
and energies $E_{\lambda} = \sqrt{(\varepsilon_{\lambda} -\mu )^{2}
+\Delta_{\lambda}^{2}}$ and solve  the set of the QRPA equations
(14) instead of one RPA equation.

The success of the single-quasiparticle approximation in describing
the magnetic moments of semi-magic nuclei makes it of interest to
try to use the same approach for quadrupole moments. In this
article, we do such analysis limiting ourselves with odd neighbors
of the even tin and lead isotopes considered in the previous
Section. To our knowledge, there is no systematic calculations of
quadrupole moments of these  nuclei.

 The static quadrupole moment of an odd nucleus in the single particle state
 $\lambda$  can be found in terms of the effective field (\ref{Vef_s}) with
the static external field  $ V_0 = \sqrt{16\pi /5} r^2 Y_{20}
(\theta)$ as follows \cite{AB,solov}:
 \beq\label{Qlam}
 Q^{p,n}_{\lambda} =  (u^2_{\lambda}-v^2_{\lambda}) V^{p,n}_{\lambda},
 \eeq
 where $u_{\lambda}$, $v_{\lambda}$ are the Bogolyubov coefficients and
 \beq\label{Vlam}
 V_{\lambda} =  -\frac{2j-1}{2j+2} \int  V(r) R_{nlj}^2(r)
 r^{2}dr.
 \eeq
The $j$-dependent factor in (\ref{Vlam}) appears due to the angular
integral \cite{BM1}. For $j>1/2$ it is always negative. For odd
neighbors of a magic nucleus the ``Bogolyubov'' factor in
(\ref{Qlam}) reduces to 1 for a particle state and to $-1$ for a
hole one.

 Components of the static effective field ${\hat V}(\omega{=}0)$, that is
$V^{n,p}(r)$ and $d_n^+(r){=}d_n^{(1)}(r)+d_n^{(2)}(r)$,  are
displayed in Figs. 15, 16 for $^{204}$Pb and  $^{116}$Sn nuclei,
correspondingly. Note that the identity $d^-(\omega {=}0){=}0$ takes
place \cite{AB}. One can see large surface maxima  of the quantities
$V^{n,p}(r)$ similar to those  in Figs. 9, 12 for the BM-like
transition amplitudes $g^{(0)}_{n,p}(r)$. In-volume (``quantum'')
corrections are relatively small, therefore the integral in Eq.
(\ref{Vlam}) is always positive. For protons, it is noticeably
larger than the similar integral with the bare field $V^{0}$, see
the  discussion on the effective charges below.

Diagonal matrix elements (\ref{Vlam}) of the proton effective field
are displayed in Fig. 17 for the tin isotopes and in  Fig. 18, for
the lead ones. As it is seen, for a  major part of the tin isotopes,
the difference between values of proton matrix elements
$V^p_{\lambda}$ surface and volume pairing is quite small. Only for
$^{112-116}$Sn nuclei it reaches 10\%. In the lead region,  the
difference is more pronounced reaching $\simeq 30\div 40$\% for
$9/2^-$ and $11/2^-$ states.

Corresponding quadrupole moments for nuclei with odd proton number
$Z=50\pm 1$ and $Z=82\pm 1$  are presented in Table III.  As it was
noted above, in this case the Bogolyubov factor in (\ref{Qlam}) is
trivial, equal to $\pm 1$. In order to check our approach, we
selected only nuclei where there are experimental data
 and those which satisfy presumably the single-quasiparticle
 approximation. In particular, we excluded several light Tl isotopes
 with known quadrupole moments of low-lying excited $9/2^-$ states. If to suppose that they
 are single-quasiparticle $1h_{9/2}$ states, they should have essentially higher
 excitation energies than it takes place.

\begin{table}[t]
\caption{Quadrupole moments $Q\;$(e\;b) of odd-proton nuclei.}

\begin{tabular}{l c c c c c c c}
\hline \hline

nucl.  &$\lambda_0$  & $Q_{\rm exp}$ &\hspace*{1.ex} $Q_{\rm
th}^{\rm surf}$\hspace*{1.ex} &\hspace*{1.ex}$Q_{\rm th}^{\rm
vol}$\hspace*{1.ex} &\hspace*{1.5ex} $Q_0$\hspace*{1.5ex} &$e_{\rm
eff}^{\rm surf}$
&$e_{\rm eff}^{\rm vol}$\\
\hline

$^{105}$In & $1g_{9/2}$& +0.83(5)&+0.83  &+0.90  &+0.18& 4.6  &   5.0 \\

$^{107}$In & $1g_{9/2}$& +0.81(5)&+0.98  &+1.07  &+0.18& 5.4  &   5.9 \\

$^{109}$In & $1g_{9/2}$& +0.84(3)&+1.11  &+1.14  &+0.18& 6.2  &   6.3 \\

$^{111}$In & $1g_{9/2}$& +0.80(2)&+1.16  &+1.10  &+0.19& 6.1  &   5.8 \\

$^{113}$In & $1g_{9/2}$& +0.80(4)&+1.12  &+1.02  &+0.19& 5.9  &   5.4 \\

$^{115}$In & $1g_{9/2}$& +0.81(5), 0.58(9) &+1.03  &+0.97  &+0.19& 5.4  &   5.1 \\

$^{117}$In & $1g_{9/2}$& +0.829(10)&+0.96  &+0.95  &+0.19& 5.1  &   5.0 \\

$^{119}$In & $1g_{9/2}$& +0.854(7)&+0.91  &+0.92  &+0.19& 4.8  &   4.8 \\

$^{121}$In & $1g_{9/2}$& +0.814(11)&+0.83  &+0.84  &+0.19& 4.4  &   4.4 \\

$^{123}$In & $1g_{9/2}$& +0.757(9)&+0.74  &+0.74  &+0.19 & 3.9  &   3.9 \\

$^{125}$In & $1g_{9/2}$& +0.71(4)&+0.66  &+0.74   &+0.19 & 3.8  &   3.9 \\

$^{127}$In & $1g_{9/2}$& +0.59(3)&+0.55  &+0.49  &+0.19& 2.9  &   2.6 \\

$^{115}$Sb & $2d_{5/2}$& -0.36(6)&-0.88  &-0.81  &-0.14& 6.3  &   5.8 \\

$^{117}$Sb & $2d_{5/2}$& -0(2)&-0.82  &-0.77  &-0.14& 5.9  &   5.5 \\

$^{119}$Sb & $2d_{5/2}$& -0.37(6)&-0.77  &-0.76  &-0.14& 5.5  &   5.4 \\

$^{121}$Sb & $2d_{5/2}$& -0.36(4), -0.45(3)  &-0.72  &-0.73  &-0.14& 5.1  &   5.2 \\

            & $1g_{7/2}^{\,*}$& -0.48(5)&-0.81  &-0.81  &-0.17& 4.8  &   4.8 \\

$^{123}$Sb  & $1g_{7/2}$      & -0.49(5)&-0.74 &-0.74     &-0.17& 4.4  &   4.4 \\

$^{205}$Tl&$3d_{3/2}^{\,*}$& 0.74(15)&+0.23 &+0.23&+0.12&1.9&1.9\\

$^{203}$Bi  & $1h_{9/2}$& -0.68(6)& -1.32   & -0.91&-0.25 & 5.3  &   3.6 \\

$^{205}$Bi  & $1h_{9/2}$& -0.59(4)& -0.94   & -0.72 &-0.25 & 3.8  &  2.9 \\

$^{209}$Bi  & $1h_{9/2}$& -0.37(3), -0.55(1)& -0.34   & -0.34 &-0.25 & 1.4  &  1.4 \\

&&  -0.77(1), -0.40(5)        &&&&&\\

\hline \hline
\end{tabular}\label{tab_Qp}
\end{table}

\begin{table}[b]
\caption{Quadrupole moments $Q\;$(e\;b) of odd-neutron nuclei.}

\begin{tabular}{lcccccc }
\hline \hline nucleus  &$\lambda_0$  & $Q_{\rm exp}$
&\hspace*{1.5ex} $Q_{\rm th}^{\rm surf}$\hspace*{1.5ex}
&\hspace*{1.5ex}$Q_{\rm th}^{\rm
vol}$\hspace*{1.5ex} &$e_{\rm eff}^{\rm surf}$ &$e_{\rm eff}^{\rm vol}$ \\

\hline

$^{109}$Sn &$2d_{5/2}$ &+0.31(10)   & +0.25  & +0.27 & 3.5 & 3.7\\

$^{111}$Sn &$1g_{7/2}$ &+0.18(9)    &+0.05  & +0.10 & 4.0 & 3.9\\

$^{113}$Sn &$1h_{11/2}^{\,*}$ &0.41(4),  0.48(5)  &-0.78  & -0.75 & 4.4 & 4.1\\

$^{115}$Sn &$1g_{7/2}^{\,*}$ &0.26(3)    &+0.38  & +0.38 & 3.9 & 3.6\\

           &$1h_{11/2}^{\,*}$ &0.38(6)   &-0.70  & -0.67 & 4.2 & 3.8\\

$^{117}$Sn &$1h_{11/2}^{\,*}$ &-0.42(5)  &-0.59  & -0.58 & 3.9 & 3.7\\

$^{119}$Sn &$2d_{3/2}^{\,*}$ &+0.094(11),   &-0.03  & -0.02 & 3.0 & 2.9\\

&&-0.065(5), &&&&\\

&&-0.061(3)&&&&\\

           &$1h_{11/2}^{\,*}$ &0.21(2)   &-0.46  &-0.45 & 3.6 & 3.5 \\

$^{121}$Sn &$2d_{3/2}$ &-0.02(2)   &+0.06  &+0.08 & 2.9 & 2.9 \\

           &$1h_{11/2}^{\,*}$ &-0.14(3)  &-0.29  & -0.29 & 3.3 & 3.3\\

$^{123}$Sn &$1h_{11/2}$ &+0.03(4)  &-0.12  & -0.10 & 3.0 & 2.9\\

$^{125}$Sn &$1h_{11/2}$ &+0.1(2)   &+0.04  & +0.06 & 2.7 & 2.7\\

$^{191}$Pb &$1i_{13/2}^{\,*}$ &+0.085(5)  &+0.0004  &+0.10 & 5.3 & 5.9\\

$^{193}$Pb &$1i_{13/2}^{\,*}$ &+0.195(10) &+0.33   & +0.39 & 6.5 & 5.5\\

$^{195}$Pb &$1i_{13/2}^{\,*}$ &+0.306(15) &+0.69 & +0.66 & 6.6 & 5.2\\

$^{197}$Pb &$3p_{3/2}$ &-0.08(17)   &+0.19   &+0.14 & 5.2 & 3.8\\

           &$1i_{13/2}^{\,*}$ &+0.38(2)   &+0.98        &+0.78 & 6.4 & 4.6 \\

$^{199}$Pb &$3p_{3/2}$ &+0.08(9)    &+0.27   & +0.19 & 4.5 & 3.1\\

$^{201}$Pb &$2f_{5/2}$ &-0.01(4)    &+0.14   & +0.09 & 4.2 & 2.8\\

$^{203}$Pb &$2f_{5/2}$ &+0.10(5)    &+0.28   & +0.22 & 3.2 & 2.3\\

$^{205}$Pb &$2f_{5/2}$ &+0.23(4)    &+0.34   & +0.28 & 2.6 & 2.0 \\

           &$1i_{13/2}^{\,*}$& 0.30(5)    &+0.67   & +0.56 & 3.0 & 2.2 \\

$^{209}$Pb &$2g_{9/2}$ &-0.3(2) &-0.26 &-0.26 & 0.9 & 0.9 \\

\hline \hline

\end{tabular}\label{tabQn}
\end{table}

 Experimental
data are taken from the compilation \cite{Q-data}.  From several
cases of proton excited isomeric states we limit ourselves with only
two, the $1g_{7/2}^{\,*}$ state in the $^{121}$Sb and
$2d_{3/2}^{\,*}$ state in $^{205}$Tl nuclei, for which the
hypothesis on the single-quasiparticle structure seems to us more or
less safe. Again, we presented results for both the kinds of nuclear
pairing (the quantities $Q_{\rm th}^{\rm surf}$ and $Q_{\rm th}^{\rm
vol}$ for surface and volume pairing, correspondingly). In the 6-th
column of the table, the single-particle quadrupole moment is
presented which is found from Eqs. (\ref{Qlam}), (\ref{Vlam}) with
substitution $V\to V_0$. As it follows from Fig. 17, for odd-proton
neighbors of the tin isotopes, difference between values of
quadrupole moments for surface and volume pairing is quite small, in
limits of 10\%. In the lead region, see Fig. 18, the difference is
 more pronounced, but here the number of the data is very
small, only 4. In addition, only in the $^{203,205}$Bi and
$^{205}$Tl  case neutron pairing exists. For these nuclei, the
effect under discussion reaches $\simeq 30\div 40$\%.

For the long chain of twelve In isotopes agreement with the data is
quite reasonable. For five Sb isotopes (six values of the quadrupole
moment) agreement is rather poor, disagreement reaching $\simeq
50\div 100$\%. A similar situation takes place for two lighter Bi
isotopes. For the $^{209}$Bi isotope where pairing is absent
experimental data are contradictory. We think that the main reason
of existing disagreements is  neglecting  the phonon coupling
effects.

\begin{figure*}
\centerline {\includegraphics [width=130mm]{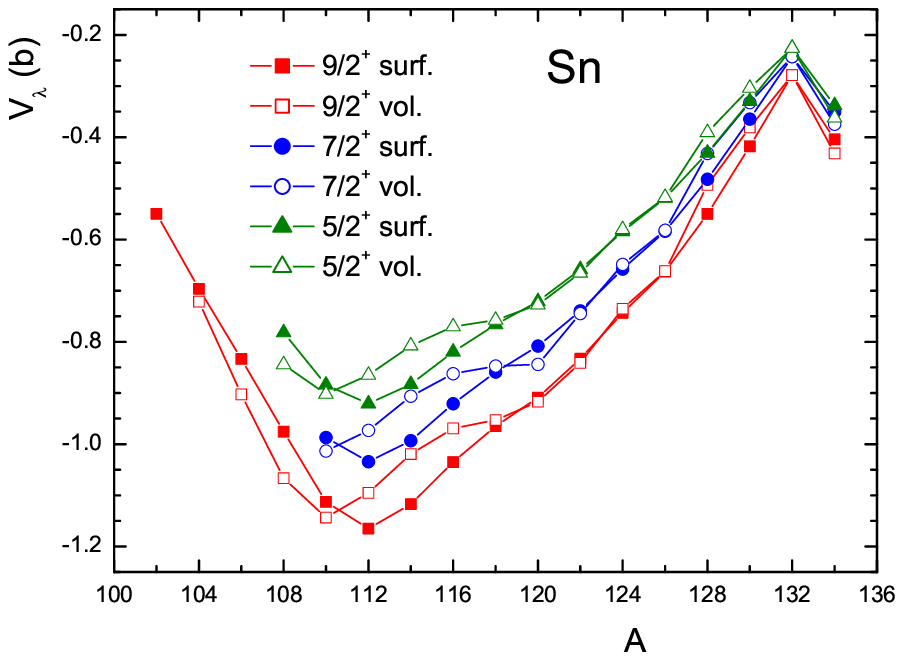}} \vspace{2mm}
\caption{(Color online) Diagonal matrix elements $V^p_{\lambda}$ of the effective
proton quadrupole field in the tin isotopes. Solid lines correspond
to surface pairing, dotted ones, to volume pairing.}
\end{figure*}

\begin{figure*}
\centerline {\includegraphics [width=130mm]{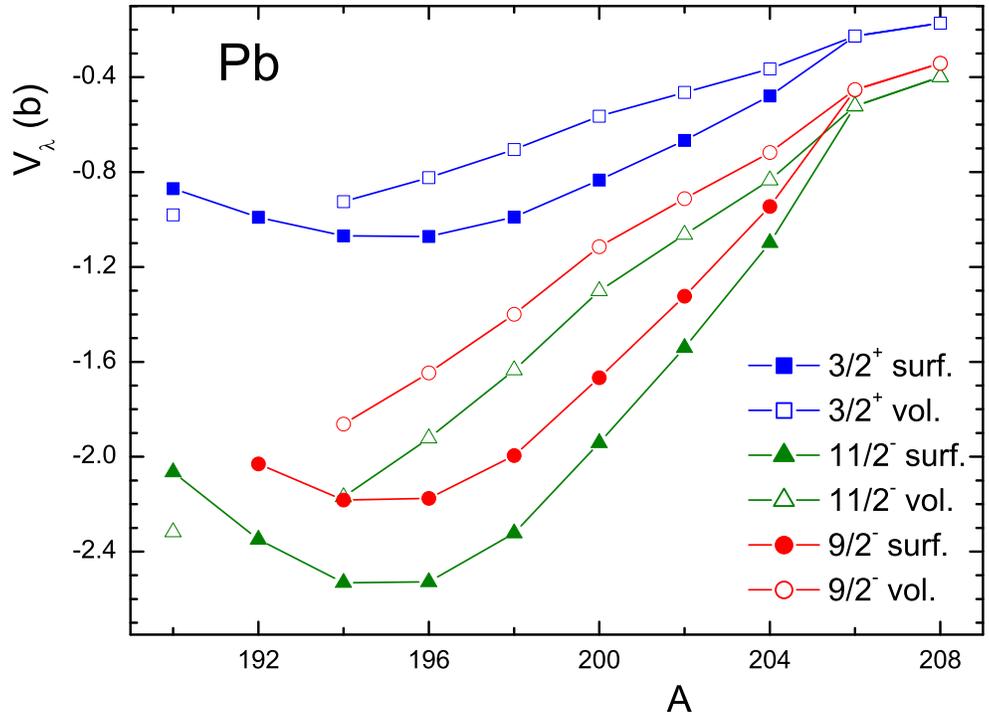}} \vspace{2mm}
\caption{(Color online) The same as in Fig. 17, but for the lead isotopes.}
\end{figure*}

Let us go to odd-neutron nuclei, the odd tin and lead isotopes. The
results are presented in Table IV and Figs. 19,20. In  selecting
nuclei for the table, we used the same concept  as for protons. In
this case, we included into the analysis twelve excited states, in
addition to the ground ones. With the only exception of the
$^{209}$Pb nucleus, all the nuclei under consideration exhibit
pairing effects and the factor $(u^2_{\lambda}-v^2_{\lambda})$ in
Eq. (\ref{Qlam}) becomes non-trivial. It changes permanently
depending on the state $\lambda$ and the nucleus under
consideration. Note that in the case of magnetic moments the factor
of $(u^2_{\lambda}+v^2_{\lambda})=1$ appears in the relation
analogous to (\ref{Qlam}) \cite{solov}. In our case, this factor
determines the sign of the quadrupole moment. In all cases when the
sign of the experimental moment is known the theoretical sign is
correct. This permits to use our predictions to determine the sign
when it is unknown.  The factor under discussion depends essentially
on values of the single-particle basis energies $\eps_{\lambda}$
reckoned from the chemical potential $\mu$ as we have
$(u^2_{\lambda}-v^2_{\lambda})=(\eps_{\lambda}-\mu)/E_{\lambda}$.
Keeping in mind such sensitivity, we found this quantity for a given
odd nucleus $(Z,N+1)$, $N$ even, with taking into account the
blocking effect in the pairing problem \cite{solov} putting the odd
neutron to the state $\lambda$ under consideration. For the
$V_{\lambda}$ value in Eq. (\ref{Qlam}) we used the half-sum of
these values in two neighboring even nuclei. We consider agreement
with the data reasonable if we have $|Q_{\rm th}-Q_{\rm
exp}|<0.1\div 0.2\;$e b. If to use such a criterion, there are 7
``bad'' cases in Table IV, and 16  ``good''. Several rather strong
disagreements with the experimental data in Table IV for high-j
levels $1h_{11/2}$ in Sn isotopes and $1i_{13/2}$ in Pb isotopes
originate just from their too distant positions from the Fermi
level. Thus, the $Q$ values depend strongly on the single-particle
level structure. Again, as for protons, the difference between
predictions of the two models under consideration is, as a rule,
rather small, and only for $1i_{13/2}$-states in the lead chain it
reaches $\simeq 20\div 30$\%.

\begin{figure*}
\centerline {\includegraphics [width=130mm]{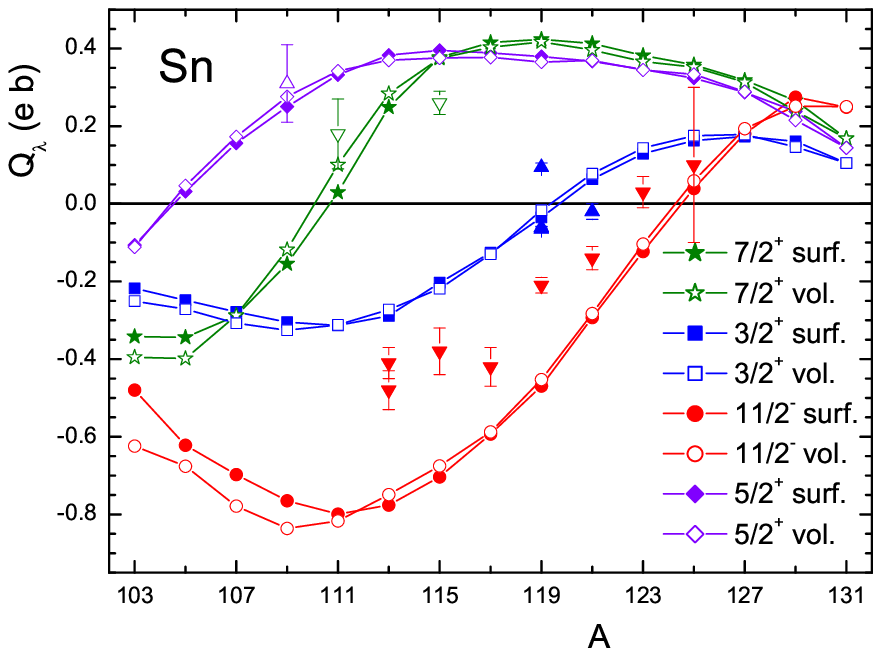}} \vspace{2mm}
\caption{(Color online) Quadrupole moments of odd tin isotopes.
Solid lines correspond to surface pairing, dotted ones, to volume
pairing. Experimental data are shown with $\blacktriangle$ for
$3/2^+$, $\blacktriangledown$ for $11/2^-$, $\vartriangle$ for
$5/2^+$, and $\triangledown$ for $7/2^+$ states.}
\end{figure*}

\begin{figure*}
\centerline {\includegraphics [width=130mm]{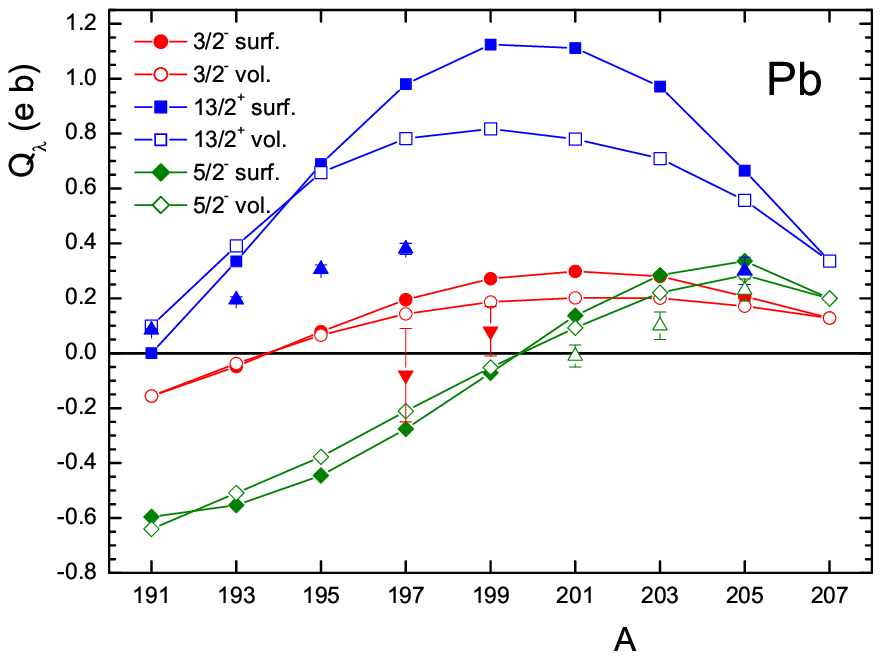}} \vspace{2mm}
\caption{(Color online) Quadrupole moments of odd lead isotopes.
Solid lines correspond to surface pairing, dotted ones, to volume
pairing. Experimental data are shown with $\blacktriangle$ for
$13/2^+$, $\blacktriangledown$ for $3/2^-$, and  $\vartriangle$ for
$5/2^-$  states.}
\end{figure*}

In the last two columns of Tables III and IV, the effective charges
are presented which are defined as $e_{\rm
eff}^{p,n}=V^{p,n}_{\lambda}/(V_0^p)_{\lambda}$. It is a direct
characteristic of the core polarizability  by  the quadrupole
external field. In these  tables, there are  only two nuclei,
$^{209}$Bi and $^{209}$Pb, with a double-magic core, and in this
case the polarizability is rather moderate, $e^{p}_{\rm eff}=1.4,
e^{n}_{\rm eff} = 0.9$. In nuclei with an unfilled neutron shell it
becomes much stronger, $e_{\rm eff}\simeq 3\div 6$. The reason is
rather obvious. Indeed, for the case of a  positive parity field
$V_0$, virtual transitions inside the unfilled shell begin to
contribute and small energy denominators appear in the propagator
${\cal L}^n$ (\ref{Ls}) playing the main role in the problem under
consideration. It enhances the neutron response to the field $V_0$
and, via the strong LM neutron-proton interaction amplitude ${\cal
F}^{np}$, the proton response as well. Results for the chain
$^{203,205,209}$Bi show how the polarizability grows with increase
of the number of neutron holes. Keeping in mind this physics, one
can represent the effective charges as  $e^{p}_{\rm eff} = 1 +
e^{p}_{\rm pol}, e^{n}_{\rm eff} = e^{n}_{\rm pol}$ where
$e^{p,n}_{\rm pol}$ is the pure polarizability charge. To separate
contributions of the unfilled shells and core nucleons explicitly,
one can divide the Hilbert space of the QRPA equations (\ref{Vef_s})
to the ``valent'' and subsidiary ones and carry out the
corresponding renormalization procedure  \cite{kaevst}.

\section{Discussion and conclusions}
The effect of the density dependence of the pairing interaction to
low-lying quadrupole excitations in spherical nuclei is analyzed
for two isotopic chains of semi-magic nuclei. Static
quadrupole moments of neighboring odd nuclei are also examined. The
complete set of the QRPA-like TFFS equations for response functions
is solved in a self-consistent way within the EDF approach to
superfluid nuclei with previously fixed parameters of the
functional. The DF3-a functional \cite{Tol-Sap} is used which is a
small modification of the functional DF3 \cite{Fay,Fay5}.
Specifically, spin-orbit and effective tensor terms of the initial
EDF DF3 were changed. Two models for effective pairing force are
considered, the surface and the volume ones, which give rise
to approximately the same accuracy in reproducing mass differences. A
noticeable effect in excitation energies $\omega_2$ is found:
predictions for the volume model are systematically higher than the
surface ones by $\delta \omega_2\simeq 200\div 300\;$keV. As to the
excitation probabilities $B(E2,{\rm up})$, the effect is not so
regular, however, as a rule, the volume values are also higher.
Thus, the correlation between these two quantities typical for the
BM model, where a higher frequency  always results in a lower
probability, is destroyed. On the average, both models reasonably
agree with the data. In addition, they both reproduce rather well
the model-independent charge density $\rho^{\rm tr}_{\rm ch}(2^+_1)$
for the  $^{118}$Sn nucleus.

Comparison with recent QRPA calculations \cite{Bertsch1} with the
Skyrme force SkM* and SLy4 shows that for the lead chain they agree
with the data a little better than our results but for the tin chain
the situation is opposite and our predictions occur to be
essentially better. The surface model is systematically better in
describing the energies $\omega_2$ whereas the excitation
probabilities are, as a rule, reproduced better with the volume
model.

Whereas the charge radii study \cite{Fay} and {\it ab initio} theory
of paring \cite{Bald1,milan3} favor  the surface pairing, the
$\omega(2^+_1)$ and $B(E2,{\rm up})$ data do not allow to prefer any
of the two kinds of pairing.

 A reasonable agreement with experiment for the quadrupole
moments of odd neighbors of the even tin and lead isotopes has been
obtained for the most part of nuclei considered. For odd-proton this
confirms that the single-quasiparticle approximation works
sufficiently well. For odd-neutron isotopes under consideration,
validity of this approach was checked previously with the analysis
of magnetic moments \cite{borzov2010}. In the case we consider, the
problem is more complicated than for odd proton isotopes as the
Bogolyubov factor
$(u^2_{\lambda}-v^2_{\lambda})=(\eps_{\lambda}-\mu)/E_{\lambda}$
comes to the quadrupole moment value, in addition to the matrix
element of the effective field $V_{\lambda}$.  This factor makes the
quadrupole moment value very sensitive to accuracy of calculating
the  single-particle energy $\eps_{\lambda}$ of the state under
consideration, especially near the Fermi surface as the quantity
$Q_{\lambda}$ vanishes at $\eps_{\lambda}=\mu$.  For such a
situation, the influence of the coupling of single-particle degrees
of freedom with phonons, see \cite{borzov2010,kaev2011}, should be
especially important. This rather complicated problem will be
considered separately.

As to the effect of the density dependence of pairing, for
quadrupole moments it is, on the average, less than for quadrupole
transitions. It depends on a nucleus examined and on the odd-nucleon
state as well. In the tin region, it is, as a rule, of the order of
$\simeq 10$\%. However, in the lead region it is higher and reaches
$\simeq 30\div 50$\% for $^{203,205}$Bi and $^{205}$Pb.

\section{Acknowledgment}
We thank J. Engel and J. Terasaki for kind supplying us with tables
of the results of the QRPA calculations \cite{Bertsch1} with the
SkM* and SLy4 force. Four of us, S. T., S. Ka., E. S., and D. V.,
are grateful to Institut f\"ur Kernphysik, Forschungszentrum
J\"ulich for hospitality. The work was partly supported by the DFG
and RFBR Grants Nos.436RUS113/994/0-1 and 09-02-91352NNIO-a, by the
Grants NSh-7235.2010.2  and 2.1.1/4540 of the Russian Ministry for
Science and Education, and by the RFBR grants 09-02-01284-a,
11-02-00467-a.

\end{document}